# Radio Bearing of Sources with Directional Antennas in Urban Environment

Cezary Ziółkowski[1], Jan M. Kelner[1]

[1] Institute of Telecommunications, Faculty of Electronics, Military University of Technology,
Gen. Witold Urbanowicz St. No. 2, 00-908 Warsaw, Poland.

*This paper focuses on assessing the limitations in the direction-finding process of radio sources with directional antennas in an urbanized environment, demonstrating how signal source antenna parameters, such as beamwidth and maximum radiation direction affect bearing accuracy in non-line-of-sight (NLOS) conditions. These evaluations are based on simulation studies, which use measurement-tested signal processing procedures. These procedures are based on a multi-elliptical propagation model, the geometry of which is related to the environment by the power delay profile or spectrum. The probability density function of the angle of arrival for different parameters of the transmitting antenna is the simulation result. This characteristic allows assessing the effect of the signal source antenna parameters on bearing error. The obtained results are the basis for practical correction bearing error and these show the possibility of improving the efficiency of the radio source location in the urbanized environment.*

Corresponding author: J. M. Kelner; email: jan.kelner@wat.edu.pl;

## I. INTRODUCTION

A multipath phenomenon is a typical feature of wave propagation in urbanized areas. This phenomenon significantly deteriorates the accuracy of direction-finding (DF) procedures, as it causes dispersion of angle of arrival (AOA) of waves at the receiver (Rx). The size of the AOA dispersion is determined by the type of propagation environment, the beamwidth and main lobe direction of the transmitting antenna pattern. Therefore, the problem of environmental impact assessment is importance in the DF procedures. In practice, different methods are used to realize this aim. These methods include signal processing techniques such as MUltiple SIgnal

Classification (MUSIC) [1–8], Estimation of Signal Parameter via Rotational Invariance Techniques (ESPRIT) [4,9–11], Space-Alternating Generalized Expectation-maximization (SAGE) [12–14], Sparse Power Angle Spectrum Estimation (SPASE) [15], and CLEAN [16]. However, none of these methods ensures minimization of the environmental impact on the DF accuracy. Knowledge of statistical properties of AOA is the basis for developing procedures to minimize bearing error. The use of measurement data for this purpose does not allow the assessment of the full range of propagation scenarios. Therefore, it is appropriate to use practically verified simulation tests.

Available literature describes only simplified ways of modelling antenna patterns, and most of these examples use the omnidirectional pattern. In this paper, the evaluation of the angular dispersion is based on simulation studies that consider the wide range of propagation phenomena occurring in the urbanized environment. The procedure provides for the generation of the set of the angles and powers of the propagation paths at Rx. This set considers the influence of the power radiation pattern of the transmitter (Tx) antenna on the AOA dispersion. This generation procedure is based on a multi-elliptical propagation model, which is presented by the authors, i.a., in [17–22]. The model maps the effects of propagation phenomena that dominate in the azimuth plane. In practice, this means that the analysis concerns transmitting antennas that the power radiation pattern is narrow in the elevation plane.

The purpose of this paper is to present the impact of the beamwidth and the main lobe direction of the transmitting antenna pattern on the angular dispersion of propagation paths at Rx, which determines the accuracy of the DF procedure. This is an important issue in the context of urbanized environments. The problem of assessing the impact of the environment on the accuracy of the DF procedure is presented by the authors in [23]. However, in that case, the analysis concerns emission sources with omni-directional antennas. Considering the directional antenna parameters and the use of statistical characteristic asymmetries of AOA for error correction is a significant extension of the presently presented assessment of environmental impact on the accuracy of the DF procedure.

The parameters of the real channels are given in the literature constitute the basis for simulation studies and analysis of the AOA dispersion. The Gaussian model [24] is used for modelling the transmitting antenna pattern in the azimuth plane. To validate the correctness of the modelling method of the angular dispersion, the comparison of the simulation and measurement results is presented for the selected measurement scenario. The obtained assessment of the impact of the antenna parameters on the AOA statistical properties is the basis for a practical procedure that ensures minimization of the impact of the propagation environment on the bearing error.

The paper structure is as follows. In Section II, the statistical properties of the bearing are presented. Section III contains the modelling procedure for determining the set of the angles and powers for the delayed and local scattering components of the received signal. The verification of this procedure, i.e., the comparison of measurement and simulation results, is presented in Section IV. For cases other than those analysed in the verification, the results of simulation studies are presented in Section V. The influence of the parameters of the transmitting antenna on the bearing error and minimizing procedure is shown in Section VI. Section VII summarizes the obtained results.

## II. STATISTICAL PROPERTIES OF BEARING

The DF process conducted in real-world conditions determines the temporary bearing, $\tilde{\theta}$, including errors, which results from the finite precision of the direction-finder and the influence of the propagation environment [23]

$$\tilde{\theta} = \theta_0 + \Delta_0 + \Delta_e = \theta_0 + \tilde{\Delta} \tag{1}$$

where $\theta_0$ is a real bearing, $\Delta_0$ is a random variable that represents the errors of the direction-finders, $\Delta_e$ is a random variable representing errors related to the propagation environment, while $\tilde{\Delta} = \Delta_0 + \Delta_e$ is the resulting random variable mapping both causes of the bearing errors.

The basic model that maps the statistical properties of the direction-finder errors is the normal distribution [25,26]. Its parameter is the RMS bearing spread, $\sigma_0$, which belongs to the set of technical parameters of the direction-finder. Depending on the sensor class, $\sigma_0$ ranges between 0.2° and 5°. The measure of the bearing error component resulting from the propagation environment is the sum of the mean values, $\bar{\varphi}_e$, and RMS angle spread (AS), $\sigma_e$. These parameters can be determined on the basis of the power azimuth spectrum (PAS) or probability density function (PDF) of angle of arrival (AOA). Thus, the resulting the DF error can be expressed as

$$\tilde{\sigma} = \sigma_0 + \bar{\varphi}_e + \sigma_e \tag{2}$$

because $\Delta_0$ and $\Delta_e$ are independent random variables.

The power balance of all propagation paths that arriving to Rx from different directions is the basis for the analytical evaluation of the statistical properties of the reception angle. Power angular spectrum (PAS), $P(\varphi_R)$, which depends on the direction of reception of individual propagation paths, is the sum of three groups of components. Based on the different times of arrival of individual paths to Rx, the PAS can be presented in the form [19]:

$$P(\varphi_R) = P_d(\varphi_R) + P_{ls}(\varphi_R) + P_{dp}\delta(\varphi_R) \tag{3}$$

where $P_d(\varphi_R)$, $P_{ls}(\varphi_R)$, and $P_{dp}\delta(\varphi_R)$ are mean PASs for the delayed scattering components, local scattering components, and direct path component, respectively, and $\delta(\cdot)$ represents the Dirac delta function.

Local extrema of measuring characteristics such as a power delay profile (PDP) or power delay spectrum (PDS) show that $P_d(\varphi_R)$ is the superposition of $N$ time-clusters

$$P_d(\varphi_R) = \sum_{i=1}^{N} P_{di}(\varphi_R) \qquad (4)$$

where $N$ is the number local extrema of PDP or PDS, i.e., the number of components received with different delays, and $P_{di}(\varphi_R)$ represents PAS of all components with $\tau_i$ delay.

But we can express PAS as the product of total power and PDF of AOA. Thus, based on (3) and (4), PDF of AOA has the form:

$$f(\varphi_R) = \sum_{i=1}^{N} \frac{P_i}{P} f_{di}(\varphi_R) + \frac{P_l}{P} f_l(\varphi_R) + \frac{P_{dp}}{P} \delta(\varphi_R) \qquad (5)$$

where $P$ is the total power at Rx, $P_i$, $P_l$, and $P_{dp}$ are the total powers of the delayed scattering components, local scattering components and the direct path component, respectively, $f_{di}(\varphi_R)$ and $f_l(\varphi_R)$ mean PDFs of AOA of the delayed and local scattering components, respectively.

The relationship between the direct component and the distributed components describes the Rice factor, $\kappa$. If the total power of components reaching Rx with a delay comparable to the carrier wave period is $P_0$ then (5) will take the form [18]:

$$f(\varphi_R) = \sum_{i=1}^{N} \frac{P_i}{P} f_{di}(\varphi_R) + \frac{1}{\kappa+1} \frac{P_0}{P} f_l(\varphi_R) + \frac{\kappa}{\kappa+1} \frac{P_0}{P} \delta(\varphi_R) \qquad (6)$$

Equation (6) is the basis for assessing statistical properties of AOA based on measurement or simulation data.

### III. MODELLING OF BEARING DISPERSION

Assessment of statistical properties of AOA based on measurement results, which in practice relate strictly to selected scenarios, limit the scope of statistical assessment of the impact of different propagation conditions. Therefore, in this paper, the global assessment of the bearing error in urbanized environments uses the results of simulation studies. The applied procedures are verified on the basis of measurement data.

The task of the simulation procedure is to generate propagation path parameters, such as AOA and power, which are the basis for determining PDF of AOA. The developed procedure uses geometric and statistical methods of modelling propagation phenomena and is based on (6).

The determination of $f_{di}(\varphi_R)$ component is basis on the Doppler multi-elliptical channel model (DMCM) [17]. The set of ellipses with foci in Tx and Rx is the geometry of this model that maps the locations of the signal scatterers. Parameters of the temporal channel characteristics such as the PDP or PDS are used to define parameters of the individual ellipses. The number and arguments of the local extremes for difference between PDP/PDS and a trend

line of these characteristics define the parameters of each ellipse

$$2a_i = D + c\tau_i, \quad e_i = D/(2a_i) \tag{7}$$

where $D$ is the distance between Tx and Rx, $\tau_i$ is the argument of the $i$th local extreme, $i = 1, 2, ..., N$, $N$ is the number of the local extremes, i.e., the number of the ellipses, c is the propagation speed of the electromagnetic wave, $2a_i$ and $e_i$ are the major axis and eccentric of the $i$th ellipse.

For the local scattering, the evaluation of $f_l(\varphi_R)$ is based on the standard PDFs such as the Laplacian [27], Gaussian [28], or logistic [29]. In this case, these properties describe the von Mises distribution [30]

$$f_l(\varphi_R) = \frac{\exp(\mu \cos \varphi_R)}{2\pi I_0(\mu)} \quad \text{for} \quad \varphi_R \in \langle -\pi, \pi) \quad \text{and} \quad \mu \geq 0 \tag{8}$$

where $I_0(\cdot)$ is the zero-order modified Bessel function and $\mu$ is a constant that represents the angular dispersion of the received signals.

Figure 1 shows the spatial geometry of the analysed problem, which considers the locations of the local and delayed scattering, the exemplary propagation path for the selected delayed scatterer, and the transmitting antenna pattern.

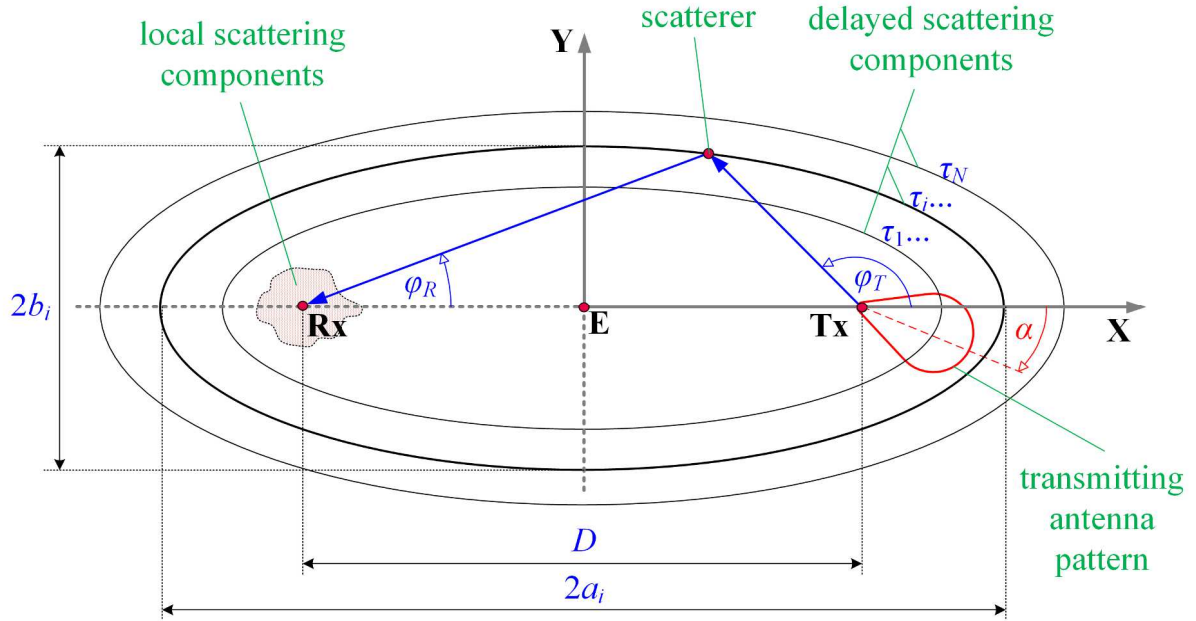

**Fig. 1.** Geometry of analysed problem.

The assumption regarding the narrow transmitting antenna pattern in the elevation plane allows one to reduce the modelling procedure to two dimensions. The following data is the basis for determining the set of angles and powers of the signal components at Rx:

- coordinate position of Tx $(x_T, y_T)$ and Rx $(x_R, y_R)$;
- PDS or PDP, which are the basis for determining the set of the delays, $\{\tau_i\}_{i=0,1,\ldots,N}$, and powers, $\{P_i\}_{i=0,1,\ldots,N}$, of the received signal components;
- number of propagation paths, $M_i$, that arrive to Rx with delay $\tau_i$;
- normalized radiation pattern of the transmitting antenna, $g_T(\varphi_T)$, and its half-power beamwidth (HPBW), $HPBW$.

The procedure for generating the angles and the powers of the delayed and local scattering components is as follows:
- determining the parameters of the ellipses, i.e., the semi-major axis, $a_i$, and eccentric, $e_i$, of the *i*th ellipse;
- generating the angles of departure (AODs), $\varphi_T$, relative to direction Tx-Rx for the propagation paths;
- determining AOAs, $\varphi_R$, for the delayed scattering components;
- for individual propagation paths (AOAs), generating the powers;
- generating AOAs for the local scattering components;
- generating the powers for the local scattering components.

The basis for AOD generation is the assumption that the occurrence probability of the scattering elements located on the ellipses with the foci in Tx and Rx is the same with respect to Tx. For the normalized radiation pattern of the transmitting antenna in the azimuth plane, we have:

$$\frac{1}{2\pi}\int_{-\pi}^{\pi} g_T^2(\varphi_T)\,d\varphi_T = 1 \quad \text{and} \quad g_T^2(\varphi_T) \geq 0 \tag{9}$$

The above relationship is the basis for defining PDF of AOD, $f_T(\varphi_T)$,

$$f_T(\varphi_T) = \frac{1}{2\pi} g_T^2(\varphi_T) \quad \text{for} \quad \varphi_T \in \langle-\pi, \pi\rangle \tag{10}$$

For the Gaussian model of the transmitting antenna pattern [24]

$$g_T(\varphi_T) = C_0 \exp\left(-\frac{\varphi_T^2}{2\sigma_T^2}\right) \quad (C_0 - \text{the normalisation constant}) \tag{11}$$

we have

$$f_T(\varphi_T) = C(\sigma_T)\exp\left(-\frac{\varphi_T^2}{\sigma_T^2}\right) \quad \text{for} \quad \varphi_T \in \langle-\pi, \pi\rangle \tag{12}$$

where $\sigma_T = \dfrac{HPBW}{2\sqrt{\ln 2}} \cong 0.6 HPBW$ and $C(\sigma_T) = \left( \int_{-\pi}^{\pi} f_T(\varphi_T) d\varphi_T \right)^{-1} = \left( \sqrt{\pi} \sigma_T \operatorname{erf}(\pi/\sigma_T) \right)^{-1}$.

Equation (12) is the basis for the generation of a set of AODs. The relationship between AOD, $\varphi_T$, and AOA, $\varphi_R$, results from the ellipse properties [19]

$$\cos \varphi_{Ri} = \frac{2e_i + \cos \varphi_T (1 + e_i^2)}{1 + e_i^2 + 2e_i \cos \varphi_T} \quad (13)$$

Hence, $\varphi_R$ is [22]

$$\varphi_{Ri} = \operatorname{sign}(\varphi_T) \cdot \arccos \frac{2e_i + \cos \varphi_T (1 + e_i^2)}{1 + e_i^2 + 2e_i \cos \varphi_T} \quad (14)$$

In the case of the local scattering, AOAs are generated directly on the basis of (8).

Each propagation path, i.e., each AOA corresponds to the random power, $p_{ij}$ (the $j$th path of the $i$th ellipse). For the delayed scattering components, the powers, $p_{ij}$, are generated on the basis of the uniform distribution

$$f(p_{ij}) = \begin{cases} M_i/(2P_i) & \text{for } p_{ij} \in \langle 0, 2P_i/M_i \rangle \\ 0 & \text{for } p_{ij} \notin \langle 0, 2P_i/M_i \rangle \end{cases} \quad (15)$$

Let us note that the mean value for this distribution corresponds to the power determined on the basis of PDP/PDS. An analogous distribution is used to determine the power of the local scattering components, $p_{0j}$,

$$f(p_{0j}) = \begin{cases} (\kappa+1) M_0 / (2P_0) & \text{for } p_{0j} \in \langle 0, 2P_0/((\kappa+1) M_0) \rangle \\ 0 & \text{for } p_{0j} \notin \langle 0, 2P_0/((\kappa+1) M_0) \rangle \end{cases} \quad (16)$$

where $M_0$ is the number of propagation paths at Rx without delay.

For each AOA, (15) and (16) provide the powers assignment for the individual propagation paths. As a result, we obtain a set $\boldsymbol{\Phi} = \{\varphi_{Rij}, p_{ij}\}_{\substack{i=0,1,\ldots,N \\ j=1,2,\ldots,M_i}}$ that represents the parameters of the propagation paths at Rx. This set is the basis for the dispersion evaluation of the reception angle as a function of $HPBW$ and the direction of the transmitting antenna pattern for the different propagation environments.

The assessment of intensity of the angular dispersion, $\sigma_e$, of the received signals is based on the PDF of AOA. From (6), it follows that the estimation of $f(\varphi_R)$ is reduced to estimate

$f_{di}(\varphi_R)$ and $f_l(\varphi_R)$. Estimators $\tilde{f}_{di}(\varphi_R)$ and $\tilde{f}_l(\varphi_R)$ are determined on the basis of $\Phi$ that is the result of simulation studies [20]

$$\tilde{f}_{di}(\varphi_R) = \frac{\sum_{j(\varphi_R)}^{M_i(\varphi_R)} p_{ij}}{\sum_{j=0}^{M_i} p_{ij}} \quad \text{and} \quad \tilde{f}_l(\varphi_R) = \frac{\sum_{j(\varphi_R)}^{M_0(\varphi_R)} p_{0j}}{\sum_{j=0}^{M_0} p_{0j}} \qquad (17)$$

where $M_i(\varphi_R)$ is the number of all propagation paths at Rx with delay $\tau_i$ and angle $\varphi_R$. From (6) and (17), it follows that the PAS estimator has the form [20]

$$\tilde{f}(\varphi_R) = \frac{\sum_{i=0}^{N} \sum_{j(\varphi_R)}^{M_i(\varphi_R)} p_{ij}}{\sum_{i=0}^{N} \sum_{j=0}^{M_i} p_{ij}} = \frac{1}{P} \sum_{i=0}^{N} \sum_{j(\varphi_R)}^{M_i(\varphi_R)} p_{ij} \qquad (18)$$

The estimator (18) provides possibilities to assess the effect of the transmitting antenna parameters on the angular dispersion of the received signals.

## IV.   VERIFICATION OF BEARING STATISTIC MODEL

### A)   Test-Bed

Verification of the bearing statistical model is carried out on the basis of empirical studies. For this purpose, we used a direction-finder dedicated for Wi-Fi signals at 2.4 and 5 GHz. Figure 2 presents this direction-finder developed at the Institute of Telecommunications of the Military University of Technology (MUT).

The direction-finder is based on two Ettus Universal Software Radio Peripheral (USRP) B210 [31] operating in the range from 70MHz to 6GHz. Each USRP B210 provides a full duplex in a system of two transmitting outputs and two receiving inputs. Its analogue-to-digital converter (ADC) enables a simultaneous analysis of signals for each receiving input with a resolution of 12 bits and bandwidth of 30.72 MHz. Hence, the direction-finder supports a circular antenna array consisting of four omnidirectional antennas. The antenna array is developed for two Wi-Fi bands, i.e., 2.4 and 5 GHz.

A dedicated software for operating the direction-finder was made using the GNU Radio [32] environment, C ++ and Python programming languages. The bearing method is based on an interferometer direction-finder idea, e.g., [33,34], and the MUSIC algorithm [1–8]. The software gives the opportunity to visualize results in the form of a bearing line on a satellite map downloaded from Google Map [35]. A GPS receiver is used to locate the positioning device on the map. The bearing error in the azimuth plane is 5.28° and 5.30° for 2.4 and 5 GHz, respectively. However, the average bearing error in the elevation plane is approximately 13.5°.

Additionally, in the case of bearing a harmonic signal source, the software allows determination of a power angular profile (PAP) in the azimuth plane. This feature was used to verify the bearing model. PAP and PAS are instantaneous and averaged characteristics describing the dispersion of the received power in an angle domain, respectively. Analogously to the dispersion in a time domain, PDP and PDS represent the instantaneous and averaged transmission characteristics of a channel, respectively.

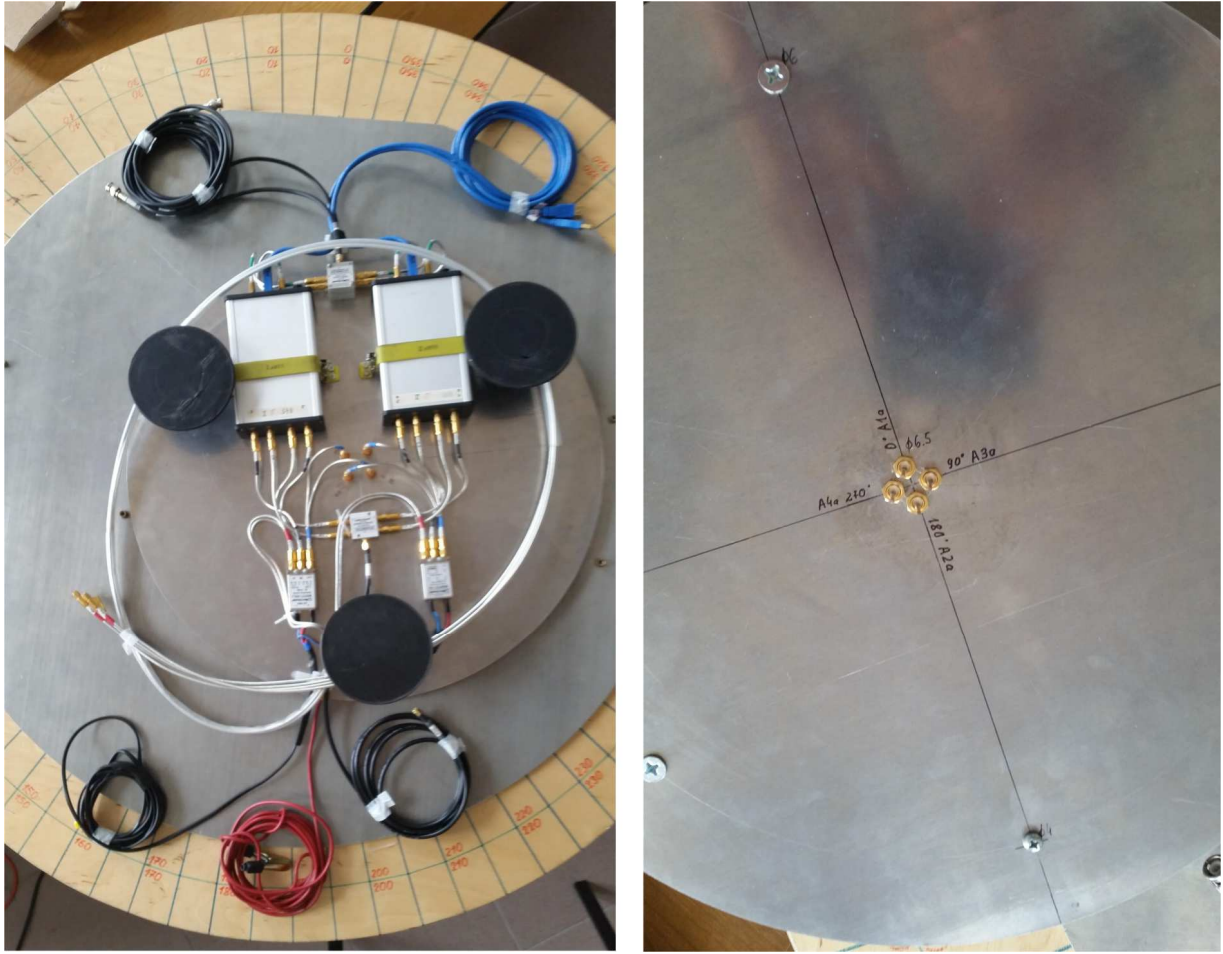

**Fig. 2.** Direction-finder based on Ettus USRP B210 with antenna array for 2.4GHz.

In the carried out empirical studies, Tx with a horn antenna was a target object. The Agilent (Keysight) E4438C ESG Vector Signal Generator [36] was used as Tx, which emitted a harmonic signal at 2.5 GHz. We used the Pasternack PE9864-10 horn antenna [37] that works at 1.7-2.6 GHz band. A gain and HPBW of the antennas are 10 dBi and 64.8°, respectively.

The basic input data for the model is PDP/PDS. Hence, sounding the radio channel was necessary. The Rohde & Schwarz FSET22 [38] was used to measure this characteristic. This broadband Rx operates in the frequency range from 100 Hz to 22 GHz. A bandwidth of a signal at the intermediate frequency (IF) is up to 500 MHz. The processing of the analogue wideband signal from the IF output is carried out by the GaGe Express CompuScope card with ADC [39]. Its maximum sampling rate is 1 GS/s.

Frequency modulated continuous wave (FMCW) signal, i.e., wideband chirp signal, at the carrier frequency of 2.5 GHz and band of 50 MHz was used to sounding the channel. In this case, the vector signal generator E4438C with an omnidirectional antenna was used as a sounding Tx.

## B) Measurement Scenario and Methodology

The comparison of the simulation results with measurement data is the verification purpose. Therefore, empirical measurements relied on determining the instantaneous PDPs and PAPs. Then, the average PDS and PASs were determined based on the PDPs and PAPs, respectively. The average channel characteristics are used to verify the model.

The empirical studies were carried out in two stages, in which PDPs and PAPs were measured, respectively. The measurements of the instantaneous characteristics were performed four times in eight measuring points located on the MUT campus. Additionally, in tests with the horn antenna, the measurement at each point was made for three selected antenna directions with respect to the Tx-Rx direction. Figure 3 depicts the Rx position and measuring points for Tx on the situational terrain map based on the Google Earth [40].

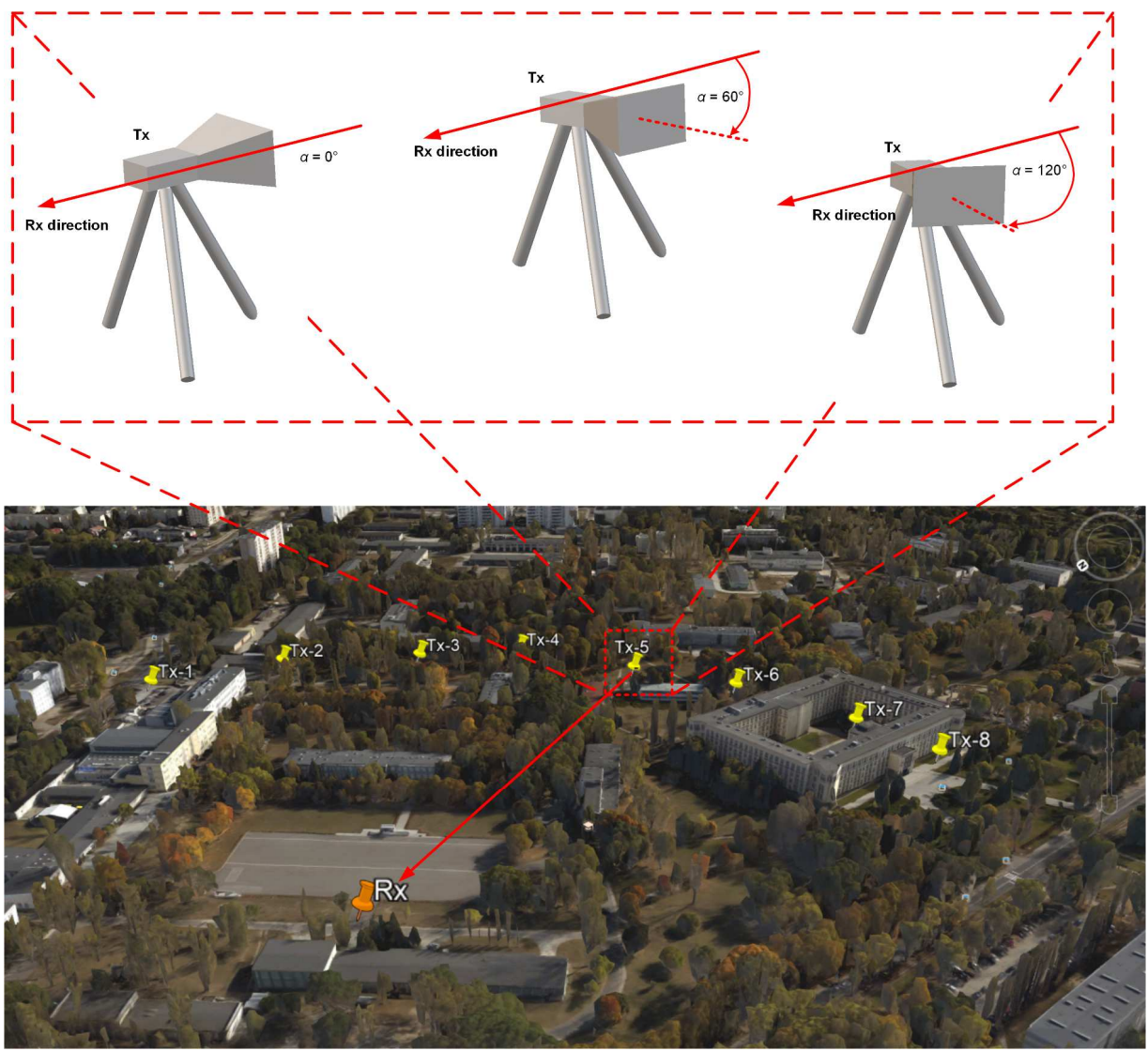

**Fig. 3.** Measurement scenario at MUT campus.

In the first stage, the R&S FSET 22 and the sounding Tx with omnidirectional antennas were located at Rx and Tx points, respectively. In the second stage, the direction-finder and generator with the horn antenna were placed at Rx and Tx points, respectively. In this case, the bearing was carried out for three fixed directions, $\alpha$, of the horn antenna, i.e., 0°, 60°, and 120°. These directions were determined in relation to the Tx-Rx direction. The three analysed antenna directions are shown in Fig. 3. The height of the transmitting and receiving antennas was 1.8 m. For all eight measurement points, the distance between Tx and Rx was approximately 300 m.

For the processing of measurement data, we applied methods described in [41,42].

## C)   Comparison of Measurement Data with Simulation Results

As a result of the first measurement stage, we obtained PDPs for four series and eight measuring points. Then, by averaging the PDPs, we determined one PDS which is shown in Fig. 4. This PDS is used in the simulation studies as the input data for the model. For the analysed environment, the RMS delay spread is 104.3 ns.

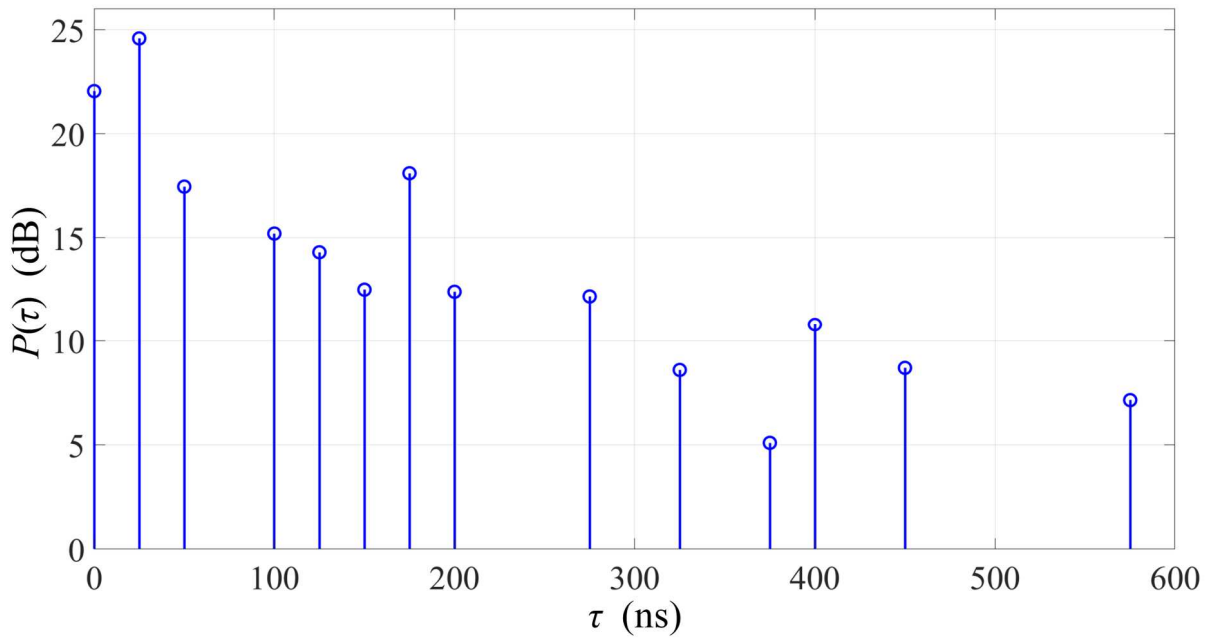

**Fig. 4.** PDS for propagation environment of MUT campus.

In the second stage of the research, we obtained the PAPs in the azimuth plane. For each direction of the horn antenna, the PAPs from all series and measurement points have been averaged. As a result, we obtained three PASs corresponding to three analysed antenna directions. In addition, using the model and empirical PDS, appropriate PASs for selected antenna directions are determined. To compare the model with the measurement data, the PASs were normalized to the PDFs of AOA. The normalized characteristics for three selected directions of the transmitter antenna are shown in Figs. 5–7, respectively.

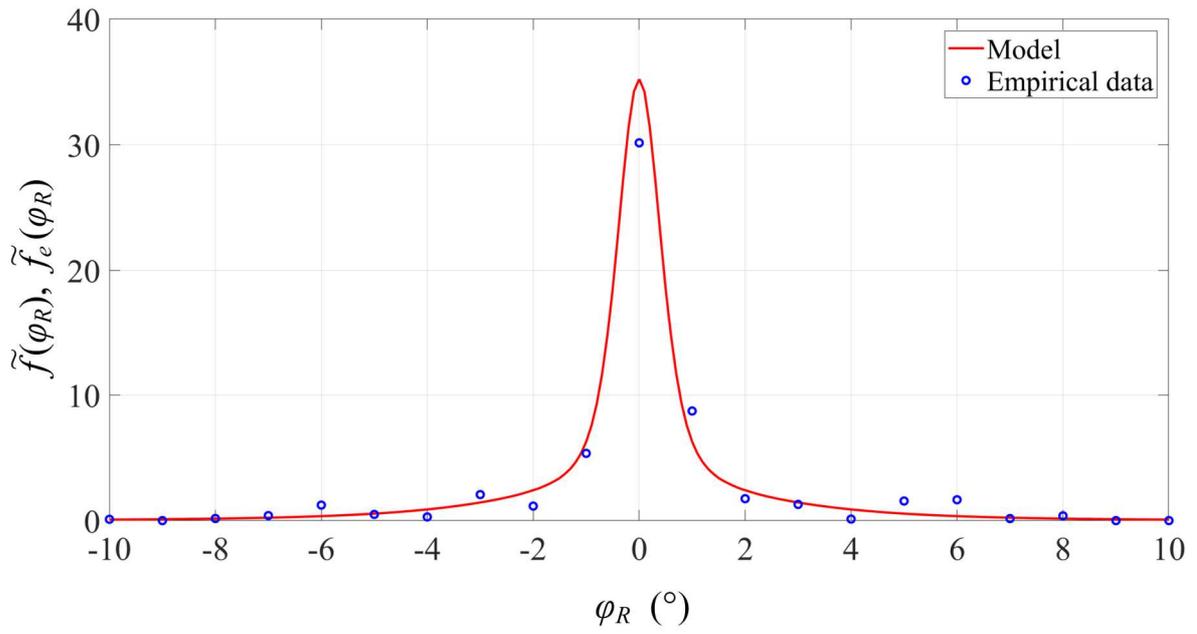

**Fig. 5.** Comparison of measurement data with model for *α* = 0°.

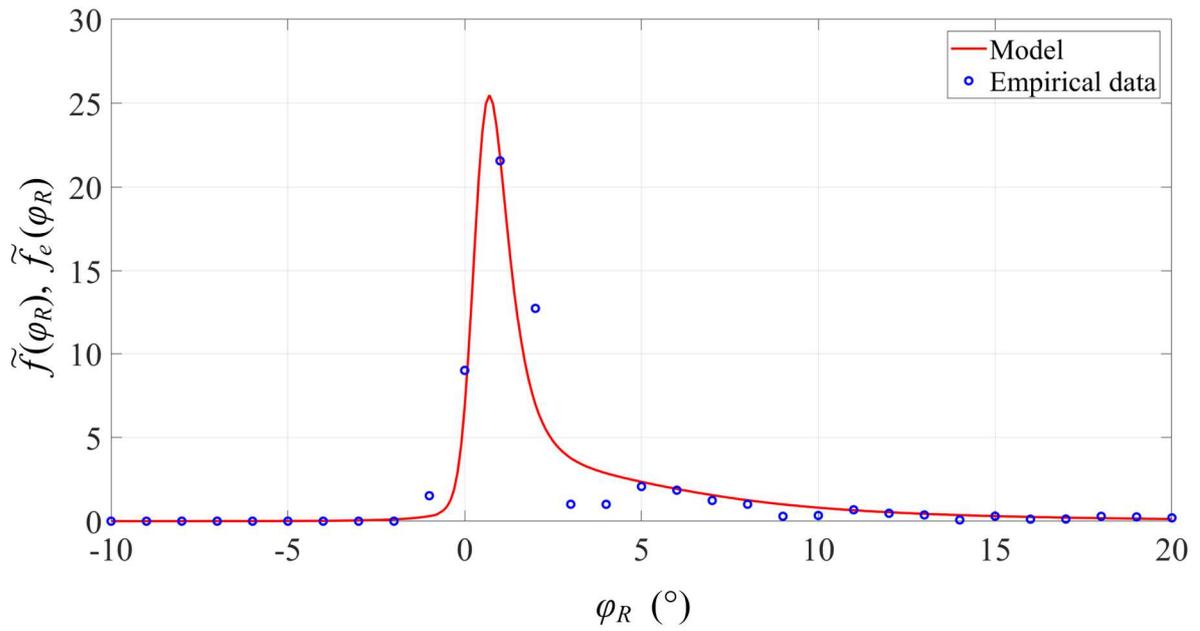

**Fig. 6.** Comparison of measurement data with model for *α* = 60°.

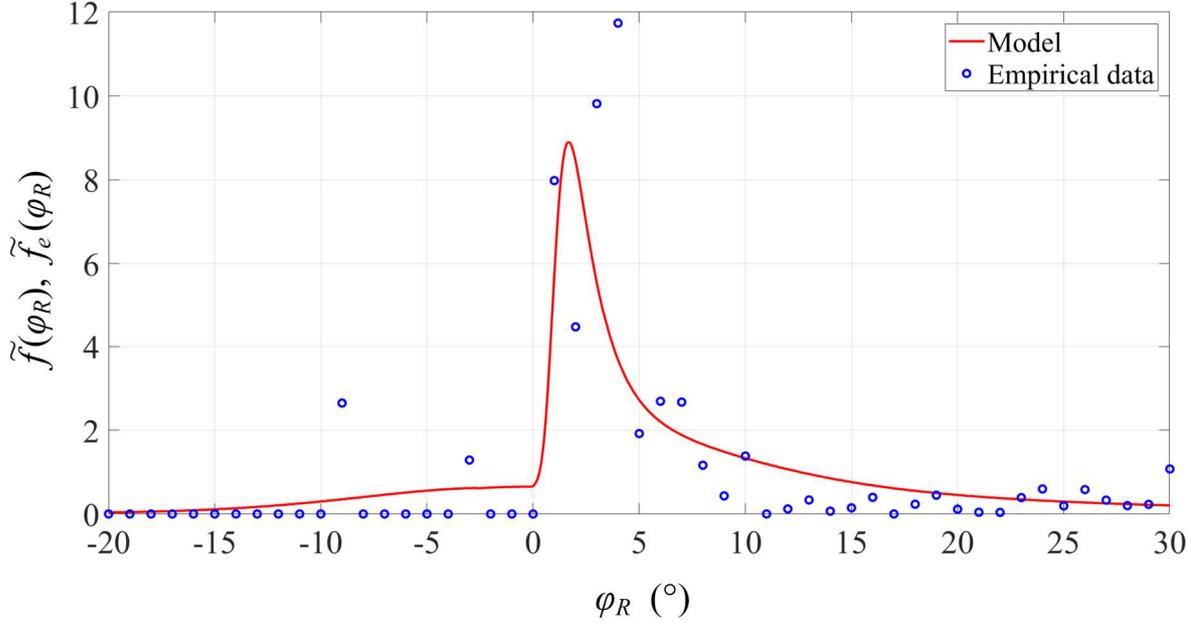

**Fig. 7.** Comparison of measurement data with model for α = 120°.

The obtained measurement results show an asymmetric nature of the angular characteristics. On the other hand, the developed statistical model of bearing well reflects the angular dispersion of environment. An approximation error of the model increases with increasing $\alpha$. To this aim, we use the least square error (LSE) defined as [26]

$$LSE = \frac{1}{K}\sum_{k=1}^{K}\left(\tilde{f}_e\left(\varphi_{Rk}\right) - \tilde{f}\left(\varphi_{Rk}\right)\right)^2 \qquad (19)$$

where $\tilde{f}_e\left(\varphi_{Rk}\right)$ and $\tilde{f}\left(\varphi_{Rk}\right)$ are the empirical and estimated PDFs of AOA obtained on the basis of measurements and (18), respectively, $\{\varphi_{Rk}\}_{k=1,2,\ldots,K}$ is the set of discreet AOAs from measurements or simulations, and $K$ is the number of all results.

In this case, we obtain $LSE$ equal to to 0.11, 0.14, and 0.34 for $\alpha = 0°, 60°, 120°$, respectively.

In remainder of the analysis, we are based on the model and empirical PDS.

## V. SIMULATION RESULTS

In this Section, we present the results of the simulation studies, which show the influence of HPBW and direction of the transmitting antenna pattern, $\alpha$, on the trend (average value) and the angular dispersion. The obtained results are an example of the assessment of the bearing error for the selected urbanized environment.

The simulation studies are performed in MATLAB on the basis of the PDS and parameters of the measurement scenario described in Section IV. In this case, we change some parameters of the signal source antenna pattern. The sets of $\mathbf{\Phi}\left(HPBW, \alpha\right)$ that consider the differentiation of both $HPBW$ and $\alpha$ are the result of these studies. The obtained data and (18) define the PDF of AOA, which is the basis for the quantitative evaluation of the dispersion of the signal reception angle. For differential HPBWs of the transmitting antenna, the PDF graphs are shown in Figs. 8 and 9 on both a linear and logarithmic scale. In this way, we show the differentiation of the results for large and small PDF values. The averaged results are obtained by a 500-fold repetition of the simulation procedure. In all simulations, we assume additional parameters: $\mu = 60$, $M_i = 60$ for $i = 0, 1, ..., N$, and $N = 13$ (see Fig. 4).

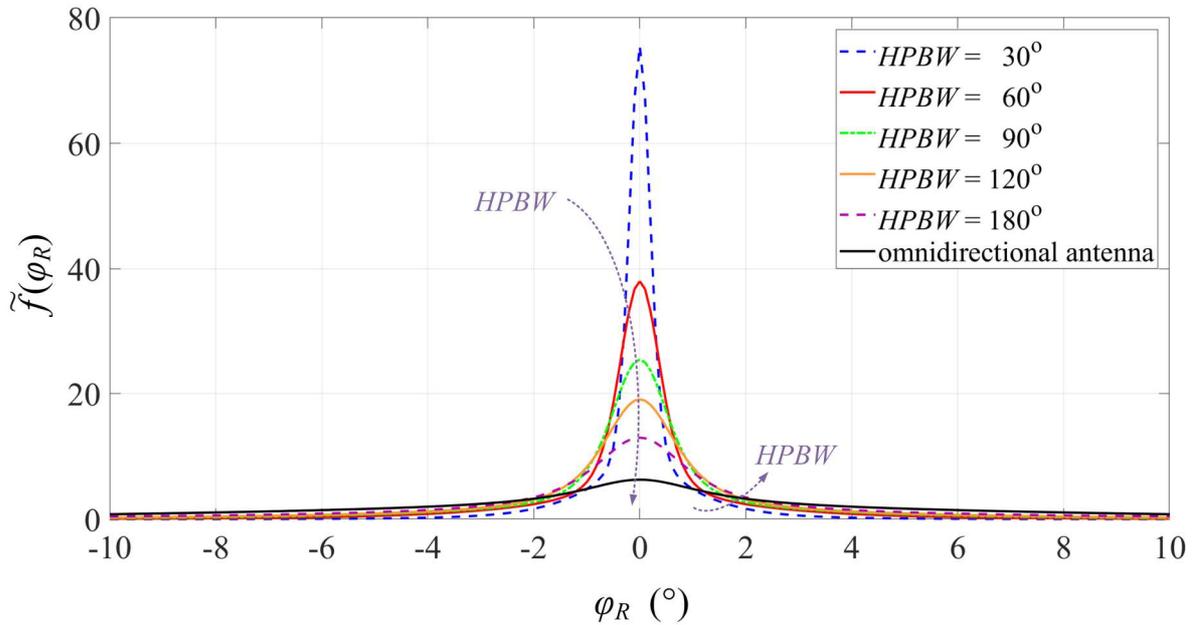

**Fig. 8.** PDFs of AOA as function of HPBW – linear scale.

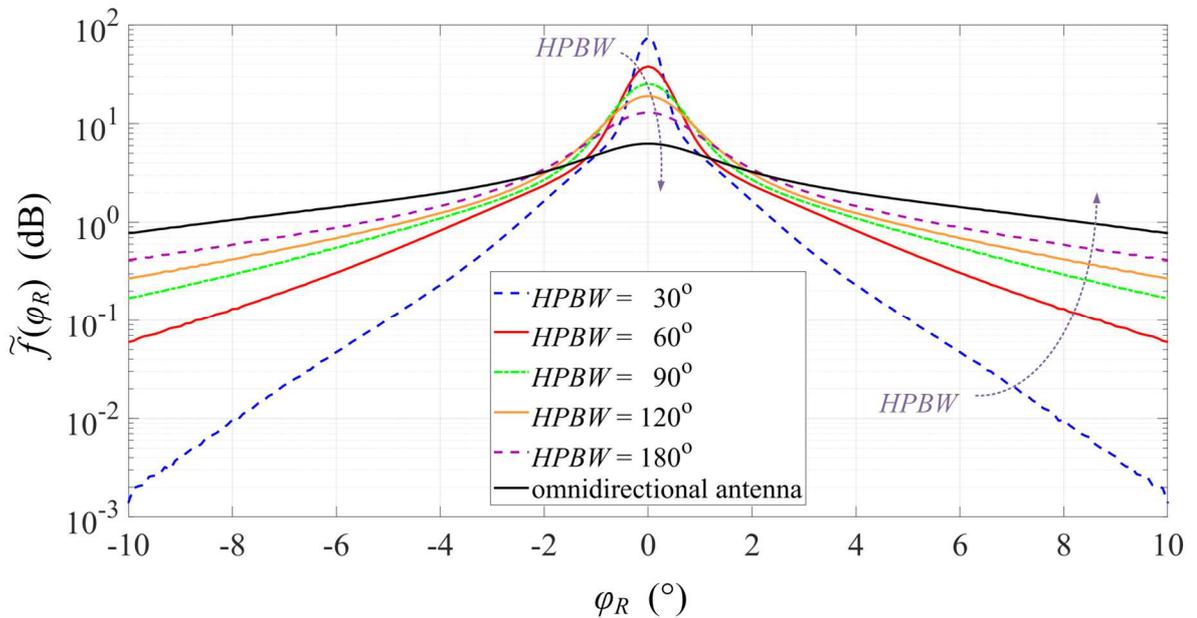

**Fig. 9.** PDFs of AOA as function of HPBW – logarithmic scale.

To assess the angular dispersion of the receive signal, AS measure is used [24]

$$\sigma_e = \sqrt{\sum_{k=1}^{K} \varphi_{Rk}^2 \tilde{f}(\varphi_{Rk}) - \left(\sum_{k=1}^{K} \varphi_{Rk} \tilde{f}(\varphi_{Rk})\right)^2} \qquad (20)$$

For the selected HPBWs, the results of the numerical calculations are presented in Table 1.

**Table 1.** Dispersion of reception angle for selected transmitting antenna patterns.

| HPBW (°) | 30 | 60 | 90 | 120 | 180 | Omnidirectional antenna |
|---|---|---|---|---|---|---|
| $\sigma_e$ (°) | 0.98 | 2.08 | 3.62 | 7.59 | 15.19 | 22.19 |

These optimistic results are only obtained when the direction of the maximum radiation of the transmitting antenna coincides with the direction determined by Tx and Rx, i.e., $\alpha = 0$. The dispersion of the reception angle depends substantially on $\alpha$. This is illustrated on the PDF graphs shown in Fig. 10. In this case, the simulation studies are performed for $HPBW = 64.8°$. Despite the sectoral radiation pattern of the transmitting antenna, the change of $\alpha$ provides not only a change in the PDF maximum but also a significant increase in the angular dispersion of the received signal. For the selected HPBWs and different $\alpha$, ASs are shown in Table 2.

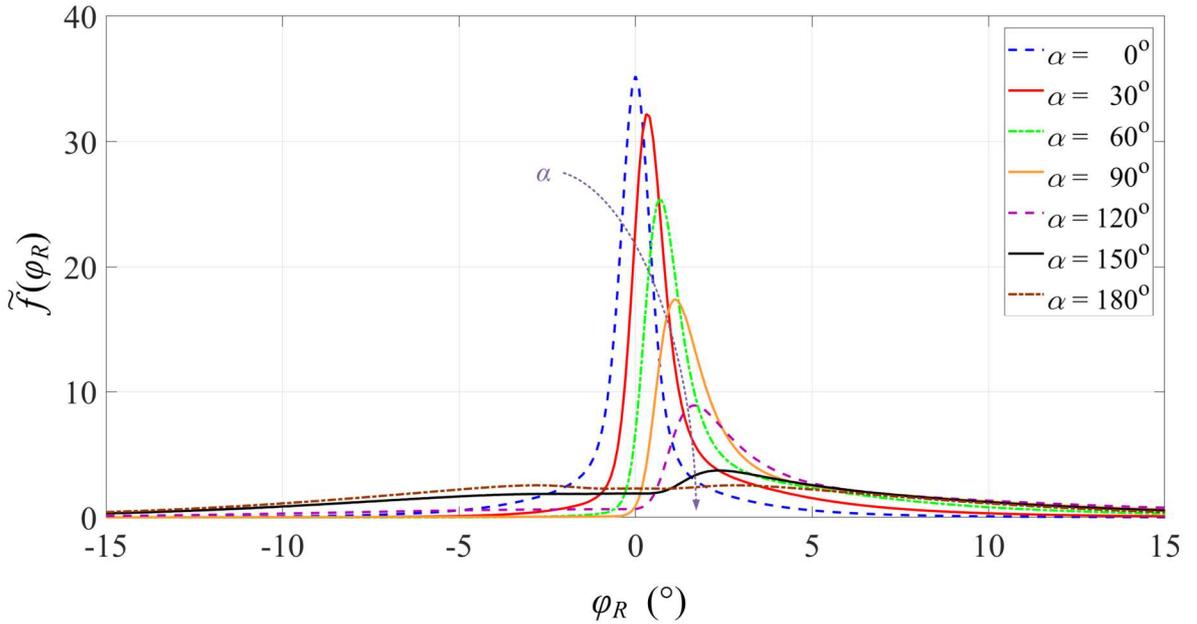

**Fig. 10.** PDFs of AOA as function of $\alpha$.

**Table 2.** Dispersion of reception angle for selected HPBWs and $\alpha$.

| | HPBW (°) | α (°) | | | | | | |
|---|---|---|---|---|---|---|---|---|
| | | 0 | 30 | 60 | 90 | 120 | 150 | 180 |
| $\sigma_e$ (°) | 30 | 0.98 | 1.78 | 3.35 | 5.75 | 10.62 | 21.12 | 13.92 |
| | 60 | 2.08 | 2.75 | 4.67 | 10.20 | 22.06 | 28.31 | 27.70 |
| | 90 | 3.62 | 5.49 | 10.43 | 18.62 | 26.52 | 29.83 | 30.33 |

| | 120 | 7.59 | 10.64 | 16.33 | 22.55 | 27.05 | 29.14 | 29.67 |

As we can see, the increase of $\alpha$ or $HPBW$ provides generally an increase of the angular dispersion. We also note the asymmetry of the PDFs, which causes the offset of the extremum of this characteristic relative to the direction to the signal source for $\alpha \neq 0$.

## VI. EVALUATION AND MINIMIZING OF BEARING ERROR

The presented modelling procedure of propagation path parameters allows to assess the effect of the radiation pattern parameters and environment on the bearing error. This error depends not only on AS, but also average offset of the reception angle, $\bar{\varphi}_e$, with respect to the signal source and reception point direction

$$\bar{\varphi}_e = \sum_{k=1}^{K} \varphi_{Rk} \tilde{f}(\varphi_{Rk}) \tag{21}$$

For $HPBW = 30°, 60°, 90°, 120°$ and different $\alpha$, the values of $\bar{\varphi}_e$ are presented in Table 3.

**Table 3.** Average offset of reception angle with respect to Tx-Rx direction.

| | HPBW (°) | $\alpha$ (°) | | | | | | |
|---|---|---|---|---|---|---|---|---|
| | | 0 | 30 | 60 | 90 | 120 | 150 | 180 |
| $\bar{\varphi}_e$ (°) | 30 | 0.00 | 1.37 | 2.95 | 5.16 | 9.26 | 5.91 | 0.00 |
| | 60 | 0.00 | 1.46 | 3.23 | 5.94 | 8.48 | 4.15 | 0.00 |
| | 90 | 0.00 | 1.72 | 3.74 | 5.71 | 5.44 | 2.81 | 0.00 |
| | 120 | 0.00 | 1.93 | 3.57 | 4.29 | 3.54 | 1.87 | 0.00 |

As we can see, an increase in beamwidth of the antenna pattern causes a significant increase in extreme $\bar{\varphi}_e$. For selected $HPBW$, the graphs of $\sigma_e$ and $\bar{\varphi}_e$ versus $\alpha$ are presented in Figs. 11 and 12, respectively.

As we can see, changing the direction of the maximum radiation of antenna causes an increase in $\sigma_e$ from a few to several degrees and $\bar{\varphi}_e$ increases from zero to a few degrees.

The dispersion analysis of the reception angle indicates that the parameters of the signal source antenna, in particular included the beamwidth and direction of maximum radiation, have a significant impact on the accuracy of the bearing determination. This means that the resulting bearing error (2) is described according to

$$\tilde{\sigma}(HPBW, \alpha) = \sigma_0 + \bar{\varphi}_e(HPBW, \alpha) + \sigma_e(HPBW, \alpha) \tag{22}$$

For selected beamwidth ($HPBW = 30°, 120°$) and class of the direction-finder ($\sigma_0 = 0.2°, 5°$), the dispersion of the source bearing versus $\alpha$ is shown in Fig. 13.

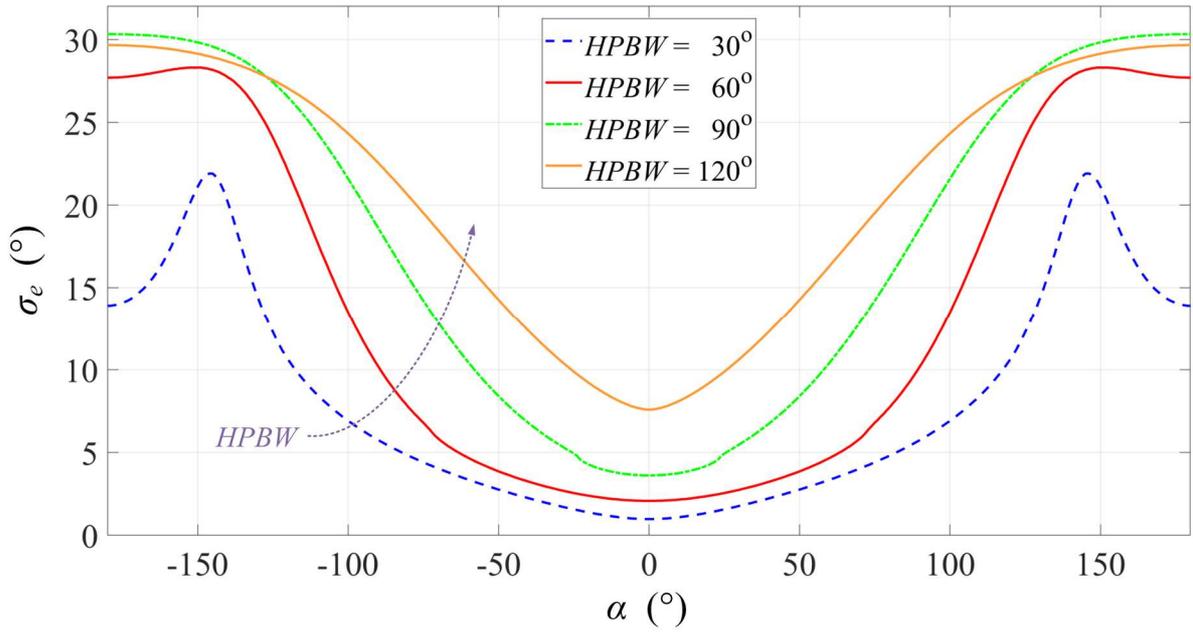

**Fig. 11.** AS versus α for selected HPBWs.

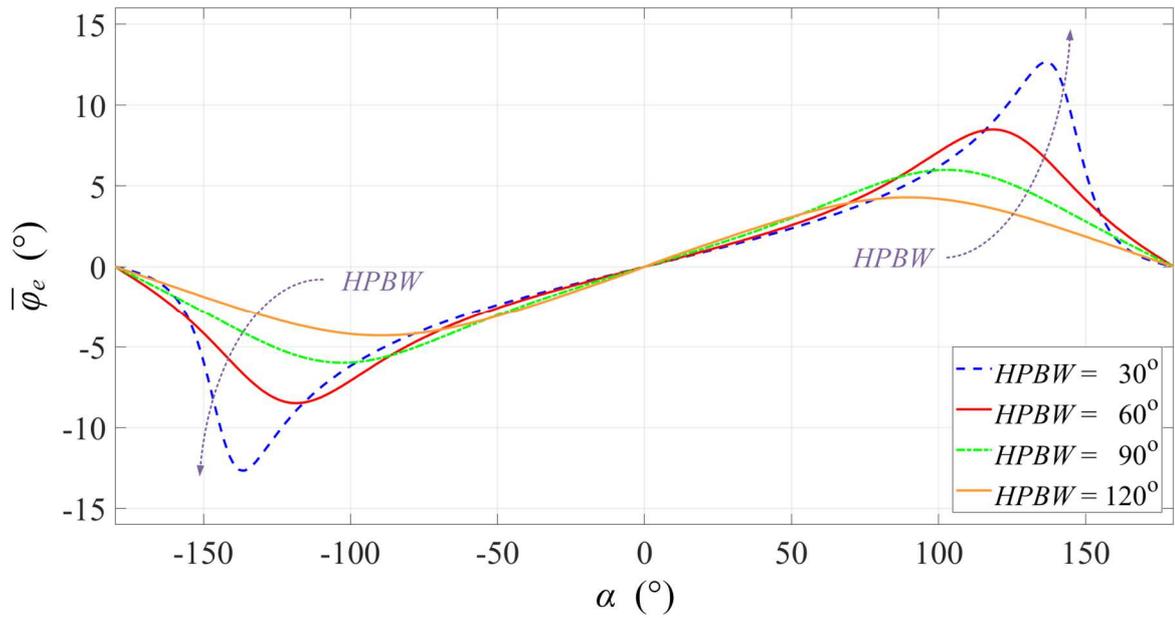

**Fig. 12.** Average reception angle versus α for selected HPBWs.

The assessment of the influence of antenna parameters on the bearing dispersion of signal source is shown on the example of simulation studies. As we can see, for the greater angular selectivity of the antenna, $\alpha$ change causes the greater range of $\tilde{\sigma}(HPBW, \alpha)$ changes and the extreme value is shifted to about $\alpha = 140°$.

The percentage evaluation of the influence of antenna parameters on the accuracy of signal source bearing indicates the significance of the problem being analysed. In this case, the percentage measure is

$$\delta(HPBW,\alpha) = \frac{\left|\overline{\varphi}_e(HPBW,\alpha)\right| + \sigma_e(HPBW,\alpha)}{\tilde{\sigma}(HPBW,\alpha)} \cdot 100\% \qquad (23)$$

For $HPBW = 30°, 120°$ and $\sigma_0 = 0.2°, 5°$, the graphs of $\delta(HPBW,\alpha)$ versus $\alpha$ are presented in Fig. 14.

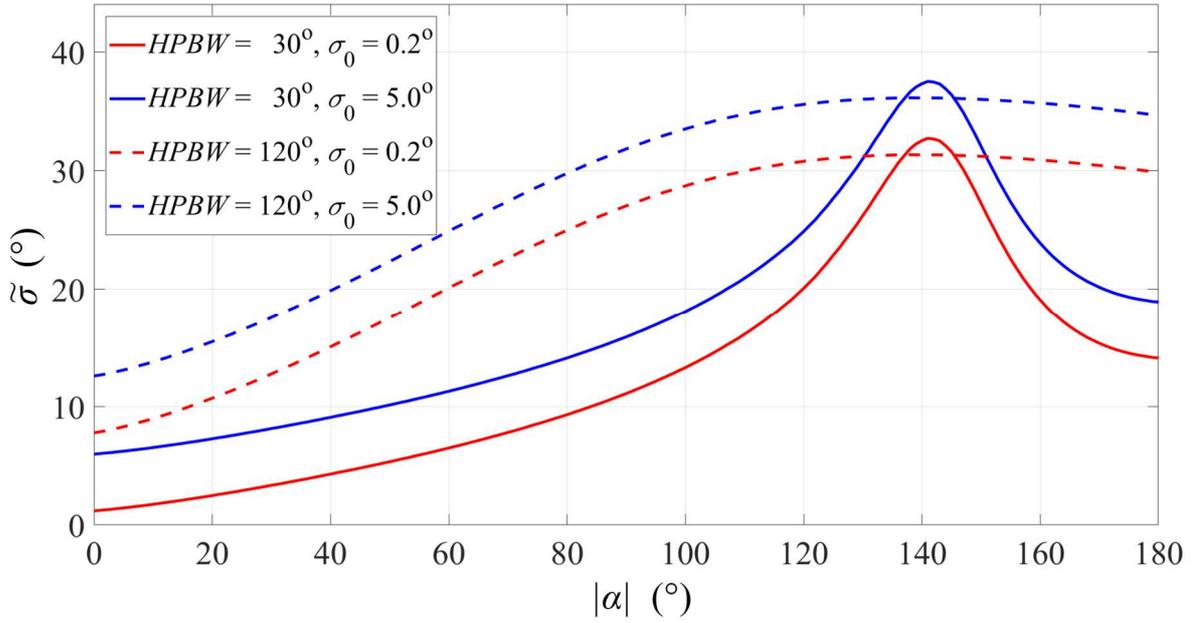

**Fig. 13.** Resulting bearing error versus $\alpha$ for selected HPBWs.

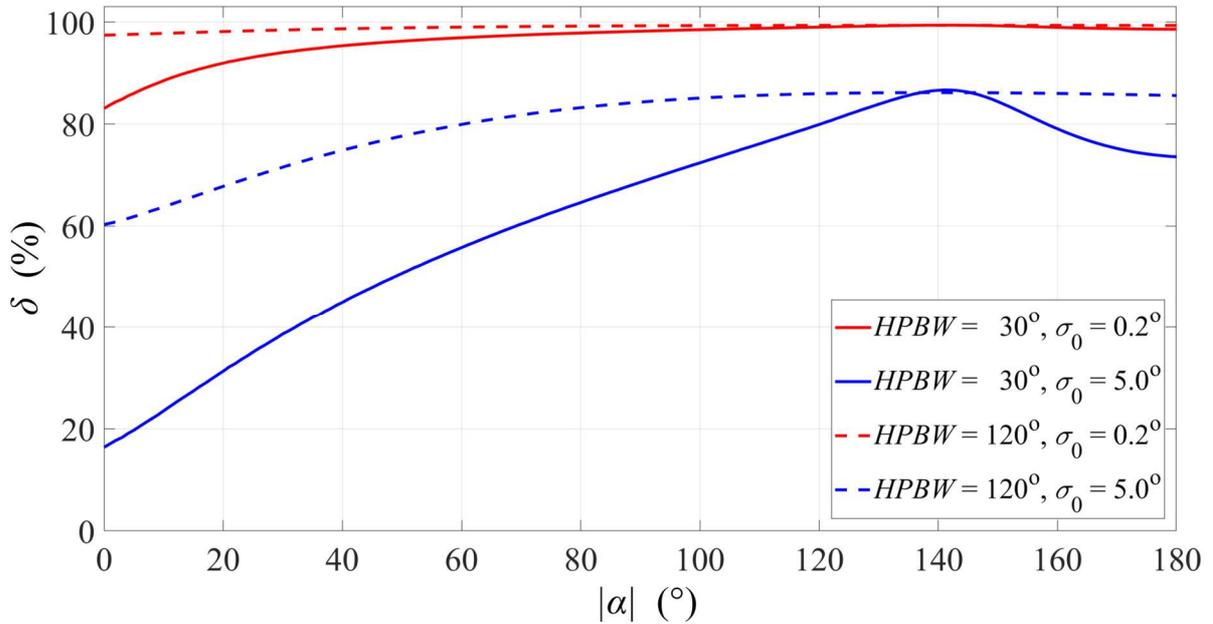

**Fig. 14.** Percentage evaluation of influence of antenna parameters on bearing error.

For direction-finder with a small bearing error, the direction of the antenna source radiation relative the position of the direction-finder has a dominant influence on the resulting bearing error. Even if the error is of the order 5°, the lack of matching of the maximum radiation direction relative to the signal source – direction-finder direction causes the error that exceeds 50% of the resulting error.

The assessment of the resulting bearing error shows importance of a search to solve the minimizing the impact of the propagation environment on the DF. We propose a solution to this problem. It is based on two assumptions. Firstly, the direction-finder can determining a PAP/PAS. Secondly, the bearing lines are determined for the directions with the highest received power. In this case, we can estimate the error of determining the bearing line, $\Delta\varphi$. This error is the difference between the actual Tx-Rx direction and the direction associated with the bearing line, i.e., the angle corresponding the maximum of the analysed PAP/PAS/PDF of AOA. Changes $\Delta\varphi$ versus $\alpha$ for different HPBWs are shown in Fig. 15.

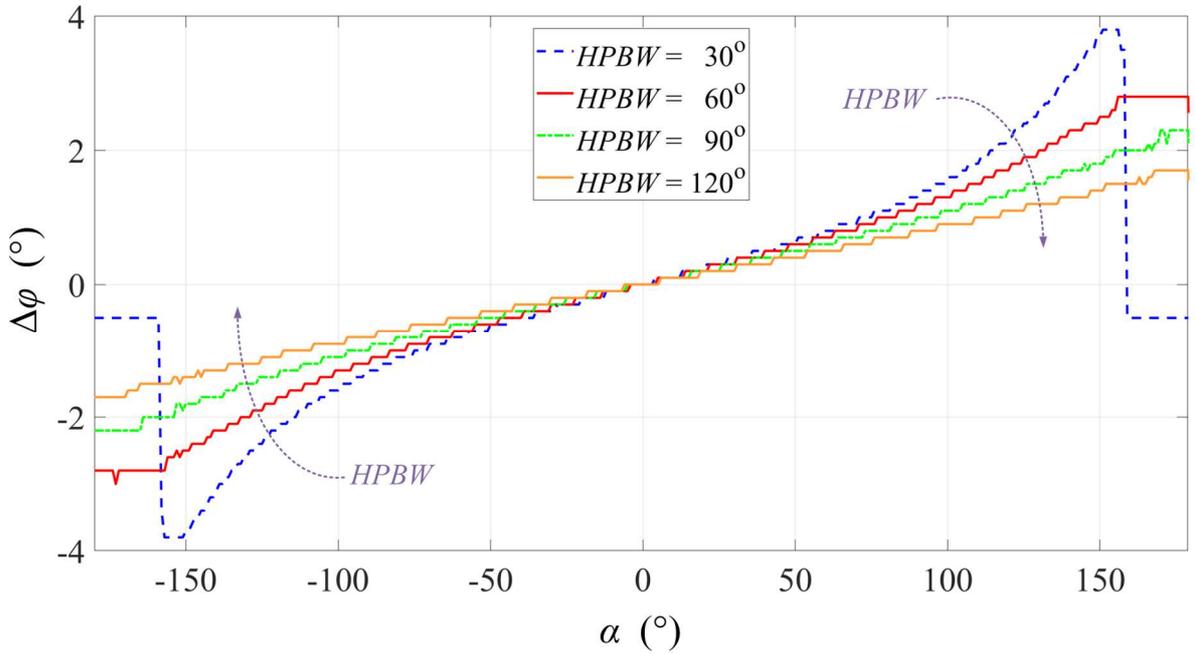

Fig. 15. $\Delta\varphi$ versus $\alpha$ for selected HPBWs.

If we compare the changes of $\Delta\varphi$ and $\bar{\varphi}_e$ versus $\alpha$ shown in Figs. 15 and 12, respectively, we can see similar nature of these changes. In the large $\alpha$ range, changes of $\Delta\varphi$ and $\bar{\varphi}_e$ are almost linear. In addition, for $\alpha \in (-90°, 90°)$, the changes of $\bar{\varphi}_e$ almost do not depend on $HPBW$. Hence, our proposal is based on the relationship between $\Delta\varphi$ and $\bar{\varphi}_e$. However, the range of variation of these parameters should be limited to the range of $\alpha$ corresponding to the extremes $\bar{\varphi}_e$ (see in Fig. 12). Then, a strong linear correlation between $\Delta\varphi$ and $\bar{\varphi}_e$ occurs. This fact is illustrated in Fig. 16.

For individual HPBWs, the presented data can be approximated with straight lines. To this aim, we use the method of least squares [43]. The obtained gradients of these lines are 0.24, 0.20, 0.17, and 0.15 for $HPBW = 30°, 60°, 90°$, and $120°$, respectively. In this case, the variation range of $\bar{\varphi}_e$ is greater for smaller HPBWs. The line gradients are similar, so we can determined a common regression line. This line with gradient of 0.21 is shown in Fig. 16.

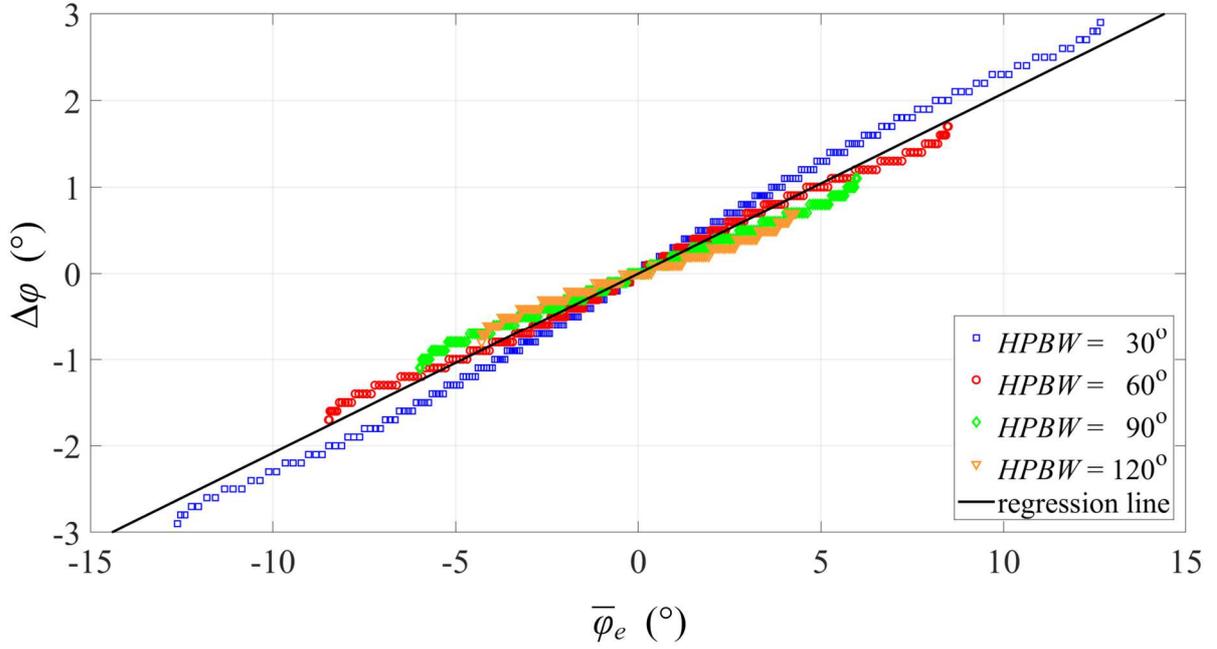

**Fig. 16.** Relationship between $\Delta\varphi$ and $\overline{\varphi}_e$ for selected HPBWs.

For the analysed data, we determine a correlation coefficient, $\rho$, defined as [43]

$$\rho = \frac{\sum_{k=1}^{K}\Delta\varphi_k \cdot \overline{\varphi}_{ek}}{\sqrt{\sum_{k=1}^{K}(\Delta\varphi_k)^2 \sum_{k=1}^{K}(\overline{\varphi}_{ek})^2}} \quad (24)$$

For all analysed HPBWs, the correlation coefficients for data sets of $(\Delta\varphi_k, \overline{\varphi}_{ek})$ are greater than 0.993. However, for a common straight regression independent of $HPBW$, $\rho$ is 0.986. This indicates the strong linear correlation between $\Delta\varphi$ and $\overline{\varphi}_e$.

To minimize the bearing error, we propose a correction

$$\vartheta = -0.21\overline{\varphi}_e \quad \text{for} \quad \overline{\varphi}_e \in (-15°, 15°) \quad (25)$$

This bearing correction is defined for the analysed propagation environment. Thus, the bearing error can be improved by up to 3°. Considering specificity of electronic warfare operation, it can be assumed that while DF a signal source, we know its communication system type, and thus its antenna type. If we have information about the antenna type, the bearing correction can be better estimated.

# VII. CONCLUSION

In this paper, the assessment of the effect of signal source antenna parameters on the bearing accuracy is presented. This evaluation is conducted on the basis of a simulation test and encompasses such parameters such as the beamwidth and direction of maximum radiation of the signal source antenna. The obtained results show that the direction-finder position relative to the direction of maximum radiation has a decisive influence on bearing error. The multi-elliptical radio channel model is used in the simulation testing procedure. In contrast to the models presented in literature, the developed procedure is based on a geometric structure, the parameters of which parameters are defined by PDP/PDS. Thanks to that, this model makes it possible to quantify the bearing error for different propagation scenarios. Verification of the simulation results with respect to the measured data indicates the correctness of the developed method of modelling multipath propagation effects in bearing procedure. The obtained results are the basis for practical correction bearing error and these show the possibility of improving the efficiency of the radio source location in the urbanized environment.

## ACKNOWLEDGEMENT
The authors would like to express their great appreciation to the anonymous reviewers for their valuable suggestions, which have improved the quality of the paper. We would also like to thank to Col. Zbigniew Piotrowski, Ph.D. D.Sc., for lending the direction-finder used in the measurement campaign presented in this paper.

**Bibliographies**

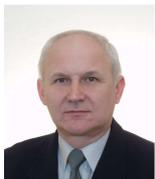
**Cezary Ziółkowski** was born in Poland in 1954. He received M.Sc. and Ph.D. degrees from the Military University of Technology (MUT), Warsaw, Poland, in 1978 and 1993, respectively, both in telecommunications engineering. In 1989 he received a M.Sc. degree from the University of Warsaw in mathematics (specialty – applied mathematical analysis). In 2013 he received a habil. degree (D.Sc.) in Radio Communications Engineering from MUT. From 1982 to 2013 he was a researcher and lecturer, and he has been a professor of Faculty of Electronics with MUT since 2013. He was engaged in many research projects, especially in the fields of radio communications systems engineering, radio waves propagations, radio communication network resources management and electromagnetic compatibility in radio communication systems. He is an author or co-author of more than 100 scientific papers and research reports.

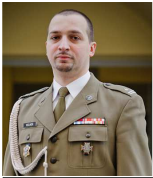 **Jan M. Kelner** was born in Bystrzyca Kłodzka, Poland in 1977. He received his M.Sc. degree in Applied Physics in 2001, his Ph.D. in Telecommunications in 2011, all from the Military University of Technology (MUT) in Warsaw, Poland. In 2011 he won "The Winner Takes All" contest on research grant of MUT Rector, and his Ph.D. thesis won the third prize in the Mazovia Innovator contest. He has authored more than 100 research articles in peer-reviewed journals and conferences. He is a reviewer for eleven scientific journals and seven conferences. He works as an assistant professor in the Institute of Telecommunications, in the Faculty of Electronics of MUT. His current research interests include wireless communications, simulations, modelling, and measurements of channels and propagation, signal–processing, navigation and localization techniques.

## List of figures and tables